\documentstyle[prd,aps,epsfig]{revtex}
\def\e{{\rm e}}
\newcommand{\be}{\begin{equation}}
\newcommand{\ee}{\end{equation}}
\newcommand{\bea}{\begin{eqnarray}}
\newcommand{\eea}{\end{eqnarray}}

\begin{document}
%\twocolumn[\hsize\textwidth\columnwidth\hsize\csname
%@twocolumnfalse\endcsname
\draft
%\preprint{gr-qc/9804056,\hspace{0.2cm} BUTP-98/12}
%\date{\today}
\date{April 22, 1998}
%\date{April 28, 1998}
\title{\bf\large 
On the quadratic action of the Hawking-Turok instanton}
\author{George Lavrelashvili
\footnote{On leave of absence from Tbilisi
Mathematical Institute, 380093 Tbilisi, Georgia}
}
\address{Institute for Theoretical Physics, University of Bern,
 Sidlerstrasse 5, CH-3012 Bern, Switzerland}
%\address{Email: lavrela@itp.unibe.ch}
\maketitle
\begin{abstract}
Positive definiteness of the quadratic part of the action 
of the Hawking-Turok instanton is investigated.
The Euclidean quadratic action for scalar perturbations is expressed 
in terms of a single gauge invariant quantity $q$. The mode functions 
satisfy a Schr\"odinger type equation with a potential $U$.
It is shown that the potential $U$ tends to a positive constant at 
the regular end of the instanton.
The detailed shape of $U$ depends on the initial data of the instanton, 
on parameters of the background scalar field potential $V$ and 
on a positive integer, $p$, labeling different spherical harmonics.
For certain well behaved scalar field potentials
it is proven analytically that for $p>1$  quadratic action is non-negative.
For the lowest $p=1$ (homogeneous) harmonic numerical solution of the 
Schr\"odinger equation for different scalar field potentials $V$ and 
different initial data show that in some cases the potential $U$ is 
negative in the intermediate region. We investigated the monotonously
growing potentials and a potential with a false vacuum.  
For the monotonous potentials no negative modes are found 
about the Hawking-Turok instanton. For a potential with the false vacuum 
the HT instanton is shown to have a negative mode for certain initial data.   
\end{abstract}
\pacs{PACS: 98.80.Hw, 98.80.Cq
\hspace{4.5cm} gr-qc/9804056\hspace{0.2cm} BUTP-98/12
}
%]

\section{Introduction}
 
A recent proposal \cite{HT1} by Hawking and Turok (HT)  
to use singular instantons to describe creation of an open inflationary 
universe caused a cascade of works 
(see e.g. \cite{Linde,HT2,Vilenkin,BL,Wu,Garriga,Unruh}).
Originally the HT instanton was discussed in
a model with a scalar field coupled to gravity.
Later it was argued that the inclusion of the four form field could 
lead to a solution of the cosmological constant problem \cite{HT3}.

It is {\it assumed} that the quadratic action of the HT instanton is
positive definite \cite{HT2}.
On the other hand it is known that some Euclidean solutions such as 
the bounce in a flat space-time, the Hawking-Moss instanton, and the 
Giddings-Strominger wormhole have the negative modes \cite{Coleman,TS,RS}. 

The aim of the present note is to investigate positive definiteness 
of the quadratic action for the HT instanton.
We shall analyze the Euclidean quadratic action which is obtained by 
the analytic continuation of the Lorentzian quadratic action for a 
scalar field in a closed universe recently derived in \cite{GMST}.
After the Hamiltonian reduction, the quadratic action can be expressed 
in terms of a single gauge invariant quantity $q$.
Expanding the perturbations by harmonics on the 
unit $3-$sphere the quadratic action can be represented as a sum of 
quadratic actions for different harmonics.
The mode functions satisfy a Schr\"odinger type equation,
with the potential $U$  depending on the background (instanton solution)
and harmonics number $p$.
We shall demonstrate that the potential tends to a positive constant 
at one (regular) end  of the instanton.
The detailed shape of $U$ depends on the instanton initial data, 
on parameters of the background scalar field potential,
and on the harmonic number $p$.
For $p>1$ we prove analytically that quadratic action is non-negative.
General proof is not working for the lowest $p=1$ (homogeneous)  harmonic.    
A numerical solution of the Schr\"odinger equation 
shows that in some cases potential $U$ is negative in the intermediate region.
For the monotonous (quadratic and quartic) scalar field potentials  
we found that the zero energy wave function has no nodes.
We investigated a case of a potential (with a false vacuum)
in which the Coleman-De Luccia (CD) solution \cite{CD} exist. 
We found that the HT instanton for a certain initial data  
(which correspond to ``overshooting'' the CD solution) has a negative mode.
 
The rest of the note is organized as follows.
In the next section we discuss main properties of the HT instanton.
In section 3 we consider the Euclidean quadratic action.
In section 4 we present our numerical results.
Section 5 contains the concluding remarks. 
 
\section{The Hawking-Turok singular instanton}
    
The Euclidean action for a scalar field  coupled to gravity has the form
\be  \label{act}
S_{E}= \int\Bigl( -\frac{1}{2 \kappa} R
+ {1\over 2} (\partial_{\mu}\varphi)^2 + V(\varphi)
\Bigr)\sqrt{g}d^4x \;,
\ee 
where $\kappa=8\pi G$ is the reduced Newton's constant 
and $V(\varphi)$ is a scalar field potential. 

We shall be interested in $O(4)$ invariant solutions and
metric is parameterized as
\be \label{sigmainterval}
ds^2=d\sigma^2+b^2(\sigma)d\Omega^2_{3}\;, 
\ee
where $d\Omega^2_{3}=d\tilde\psi^2+sin^2(\tilde\psi) d\Omega^2_{2}$
is the line element of the unit 3-sphere.

The field equations are 
\be\label{fieq}
\varphi''+3{b'\over b}\varphi'={\partial V\over\partial\varphi}, \quad
b''=-{\kappa\over 3}b(\varphi'^2+V(\varphi)) \;,
\ee
%and
%\be\label{beq}
%b''=-{\kappa\over 3}b(\varphi'^2+V(\varphi)) \;,
%\ee
where prime denotes $d/d\sigma$.

The HT instanton has the following initial conditions at $\sigma\to 0$ 
\cite{HT1}
\be \label{htincond}
\varphi(0)=\varphi_{0},\qquad\varphi'(0)=0,\qquad b'(0)=1 \;, 
\ee
which implies
\be\label{incond}
\varphi(\sigma)=\varphi_{0}
+{1\over 8}{\partial V\over\partial\varphi}_{|\varphi=\varphi_{0}}~\sigma^2
+O(\sigma^4), \quad
b(\sigma)=\sigma-{\kappa\over 18}V(\varphi_{0})\sigma^3+O(\sigma^5) \;, 
\ee
where $\varphi_{0}$ is some constant.
                                                                 
If one starts at $\sigma=0$ with the initial conditions Eq.~(\ref{htincond}),
the scale factor $b$ grows, reaches maximal value and decreases again
developing the second zero at some $\sigma=\sigma_f$. 
Next HT \cite{HT1} argue that as we approach $\sigma_f$ the potential terms  
become irrelevant in the field equations (\ref{fieq}): the first equation then 
implies that $\varphi'\propto b^{-3}$ and the second yields 
$b\propto (\sigma_f -\sigma)^{1/3}$ or 
%From the equations it follows that as $\sigma\to\sigma_f$
\be \label{sing}
\varphi=\varphi_f-\sqrt{2\over 3\kappa}\ln (\sigma_{f}-\sigma),
\qquad b=C (\sigma_f-\sigma)^{1/3} \;,
\ee
where $\varphi_f$ and $C$ are some constants.
This assumption is certainly true for potentials, which grow 
sufficiently slowly as $\varphi\to\infty$, namely
\be \label{condition}
V(\varphi(\sigma_{f}-\sigma))<\varphi'^2 \;.
\ee 
Note that if the potential $V$ contains an exponential $\varphi$ dependence,
as in hybrid inflation in supergravity \cite{Linde}, condition
(\ref{condition}) can be violated and the instanton can have different
from Eq.~(\ref{sing}) asymptotic behavior.
In what follows we assume potentials which respect Eq.~(\ref{condition}).
In the opposite case it is difficult to draw conclusions which are 
independent of a concrete form of a potential $V(\varphi)$.

The above discussion is true for any $\varphi_{0}$ for the monotonously
increasing potentials. If the potential has a local minimum 
(false vacuum) $\partial V/\partial\varphi$ changes sign  and under certain
conditions \cite{JS} can stop the motion of $\varphi$. If $\varphi'$
returns to zero precisely at $\sigma_{f}$ one obtains the Coleman-De Luccia
\cite{CD} solution. In that sense the CD solution is a 
special (non-singular) case of the HT instanton. 

Some equations would be convenient to write in terms 
of a conformal time $\tau$, 
%\be \label{tauinterval}
%ds^2=b^2(\tau)(d\tau^2+d\Omega^2_{3}) \;, 
%\ee
related to $\sigma$ by $d\tau=d\sigma/b$.
Close to the origin $\sigma=\e^{\tau}(1+O(\e^{2\tau}))$, 
and $\sigma\to 0$ corresponds to  $\tau\to -\infty$.

Close to the singularity (which we assume at $\tau=\tau_f$)
$d\sigma/d\tau=C (\sigma_{f}-\sigma)^{1/3}$
and $(\tau_f-\tau)={3\over 2C}(\sigma_f-\sigma)^{2/3}$
which gives
\be
\varphi=\tilde\varphi_f-\sqrt{3\over 2\kappa}\ln (\tau_{f}-\tau),
\qquad b=C\sqrt{{2C\over 3}(\tau_f-\tau)} \;.
\ee

\section{Euclidean quadratic action}

We will use the results of a well developed theory of gauge invariant 
cosmological perturbations to obtain the quadratic part of the Euclidean 
action. In \cite{GMST}, in particular, cosmological perturbations 
for the Lorentzian version of the theory defined by the action 
Eq.~(\ref{act}) for closed universe have been considered.

Since vector and tensor perturbations decouple in our model,
we will be interested only in scalar perturbations, 
which are parameterized as follows
\bea \label{perturbations}
\varphi&\to& \varphi (\eta)+ \delta\varphi \\ 
ds^2&=&a(\eta)^2[-(1+2A)d\eta^2+2B_{|i}dx^id\eta  
+((1-2\psi)\gamma_{ij}-E_{|ij})dx^idx^j] \;,  \nonumber
\eea
where $\gamma_{ij}$ is a metric on the unit $3-$sphere
and $A, B, E, \psi$ and $\delta\varphi$ are small quantities. 

After the expansion of the perturbations by harmonics on the 
unit $3-$sphere the Eq.~(B18) from \cite{GMST} implies 
that the quadratic part of the action is represented as 
\be \label{lact}
S^{(2)}_{L}=\sum_{p=1}^{p=\infty} {(p^2-4K)\pi^2\over 2}
\int\Bigl[ ({dq_{p}\over d\eta})^2         
+({\kappa\over 2}\varphi^2_{,\eta}
+\varphi_{,\eta}({1\over\varphi_{,\eta}})_{,\eta\eta}-p^2+4K) q_{p}^2
\Bigr]d\eta \;,
\ee 
where ${}_{,\eta}=d/d\eta$, $K=+1$ for a closed universe,
$p=1,2,...$ is an integer labeling different spherical harmonics 
and $q$ is a gauge invariant quantity
constructed from the perturbations\footnote{Since different harmonics 
decouple in what follows we shall omit the index $p$.}
\be
q={2a\over \kappa\varphi_{,\eta}}[\psi-{a_{,\eta}\over a}(B-E_{,\eta})] \;.
\ee 
Now we analytically continue the Lorentzian quadratic action Eq.~(\ref{lact}) 
to the Euclidean region taking $\eta=-i\tau$.
We have  $S^{(2)}_L=i {S}^{(2)}_E$ where quadratic part of the 
Euclidean action is represented as a sum of following terms
\be \label{eaction}
\tilde{S}^{(2)}_{E}= {(p^2-4K)\pi^2\over 2}
\int\Bigl[ ({dq\over d\tau})^2+({\kappa\over 2}\dot{\varphi}^2+
\dot{\varphi} ({1\over\dot{\varphi}})~\ddot{}+p^2-4K) q^2
\Bigr] d\tau \;,
\ee 
where $\dot{}=d/d\tau$.
We see that quadratic action for the $p=2$ mode vanishes.
Another problem is that for the lowest $p=1$ (homogeneous) harmonic
the overall sign of the quadratic action is ``wrong''.
We assume that analytic continuation $q\to -i q$ is performed 
while integrating over this mode (compare \cite{GPH,TS,RS}). 

The equation for the mode functions, 
which diagonalize the action (\ref{eaction}), 
has form of the Schr\"odinger equation 
\be \label{setau}
-{d^2\over d\tau^2}q+U q=E q \;,
\ee 
with a potential $U$ depending on the background fields
\be
U={\kappa\over 2}\dot{\varphi}^2+\dot{\varphi}
({1\over\dot{\varphi}})~\ddot{}+(p^2-4K) \;. 
\ee
With the help of the equations of motion for the background quantities 
this potential can be written in the form
\footnote{We denote the scale factor $a$ in the Euclidean region as $b$
and assume $K=+1$ in what follows.}
\be \label{upot}
U=4{\dot{b}^2\over b^2}+
{2b^4\over \dot{\varphi}^2}({\partial V\over\partial\varphi})^2
+(p^2-4)
-{8b\dot{b}\over\dot{\varphi}}{\partial V\over \partial\varphi}
-b^2{\partial^2V\over\partial\varphi^2}-{2\kappa\over3}b^2V
-{\kappa\over 6}\dot{\varphi}^2 \;.
\ee

One can check that close to the origin $\sigma\to 0$ 
where $\tau\to -\infty$ the first four terms contribute 
to the leading behavior and the potential tends to a positive constant 
\be
U=p^2+O(\e^{2\tau}) \;. 
\ee
Correspondingly the regular branch of the wave function 
dies out exponentially 
\be\label{regulartau}
q=N\e^{\sqrt{p^2-E}~\tau} (1+O(\e^{2\tau})) \;,
\ee
where $N$ is a normalization constant. Note that $E<p^2$
corresponds to bound states and $E>p^2$ to scattering states.

Towards the singular instanton end (as $\tau\to\tau_f$)
for sufficiently slowly growing scalar field potentials 
(condition (\ref{condition}))
only the first and the last terms dominate in Eq.~(\ref{upot}) and give 
\footnote{This fact was independently noticed in \cite{Garriga}.} 
\be \label{singpotential}
U={3\over{4(\tau_f-\tau)^2}}+O((\tau_f-\tau)^0) \;.
\ee
If the potential $V$ grows faster, $U$ can change its asymptotic behavior 
and even sign, which might lead to catastrophic quantum mechanical
behavior.

The regular branch of the wave function in the potential 
Eq.~(\ref{singpotential}) close to $\tau_f$ behaves as
\be\label{regularsing}
q\approx (\tau_f-\tau)^{3/2} \;.
\ee
Note that in the derivation of the quadratic action Eq.~(\ref{eaction})
we neglected the surface terms, which appear after integration by parts. 
The behavior Eq.~(\ref{regulartau}) and Eq.~(\ref{regularsing})
justify this procedure.
 
In the intermediate region  $-\infty<\tau<\tau_f$ the potential $U$ 
can be positive as well as negative depending on the details of the 
scalar field potential $V(\varphi)$, the initial value $\varphi_{0}$ 
and the harmonic number $p$. 

After rescaling $q=f/\dot{\varphi}$ and integrating by parts
we obtain the action 
\be \label{poseaction}
\tilde{S}^{(2)}_{E}= {(p^2-4)\pi^2\over 2}
\int\Bigl[ ({df\over d\tau})^2+({\kappa\over 2}\dot{\varphi}^2+p^2-4)f^2
\Bigr]{d\tau\over\dot{\varphi}^2} \;,
\ee
which is obviously non-negative for $p>1$. 
For $p=1$ we have to investigate the problem numerically. 

\section{Numerical results}

Let me describe our numerical results.
The main idea is to study the zero energy wave function.
Nodes of zero energy wave functions usually mean 
that there are bound states with negative energies. 
The number of nodes counts the number of bound states.

For us it was convenient to solve the equations in $\sigma$.
The $\sigma$ analogue of the Eq.~(\ref{setau}) has the form
\be\label{sesigma}
{d^2\over d\sigma^2}q=-{b'\over b}{d\over d\sigma}q
+(W-{E\over b^2})q \;,
\ee
where 
\be
W=4{b'^2\over b^2}+{2\over{\varphi'}^2}({\partial V\over\partial\varphi})^2
+{p^2-4\over b^2}-{8b'\over{b\varphi'}}{\partial V\over \partial\varphi}
-{\partial^2V\over\partial\varphi^2}-{2\kappa\over3}V
-{\kappa\over 6}{\varphi'}^2 \;.
\ee
One finds that the regular at $\sigma\to 0$ branch of the wave function 
behaves as
\be\label{regularsigma}
q=\tilde{N}\sigma^{\sqrt{p^2-E}}(1+O(\sigma^2)) \;,
\ee 
where $\tilde{N}$ is a normalization factor.
In what follows we shall use $\kappa=1$ units.

Starting at $\sigma=\epsilon$ with $\epsilon=10^{-3}\div 10^{-4}$ 
we integrated simultaneously the background equations 
(\ref{fieq})  with initial conditions (\ref{incond}) and  the
Schr\"odinger equation (\ref{sesigma}) with initial condition 
(\ref{regularsigma}) for $p=1, E=0$, with different scalar field potentials 
and different initial values $\varphi_0$. 
We investigated quadratic and quartic (monotonous) scalar field 
potentials $V$ (up to parameter values $V\approx M^4_{Pl}$) and 
a potential with the false vacuum.
The shape of the potential $U$ in the intermediate region
depends on the initial conditions and  details of the 
background scalar field potential $V(\varphi)$. 

\begin{figure}
\centering
\includegraphics[width=0.50\textwidth]{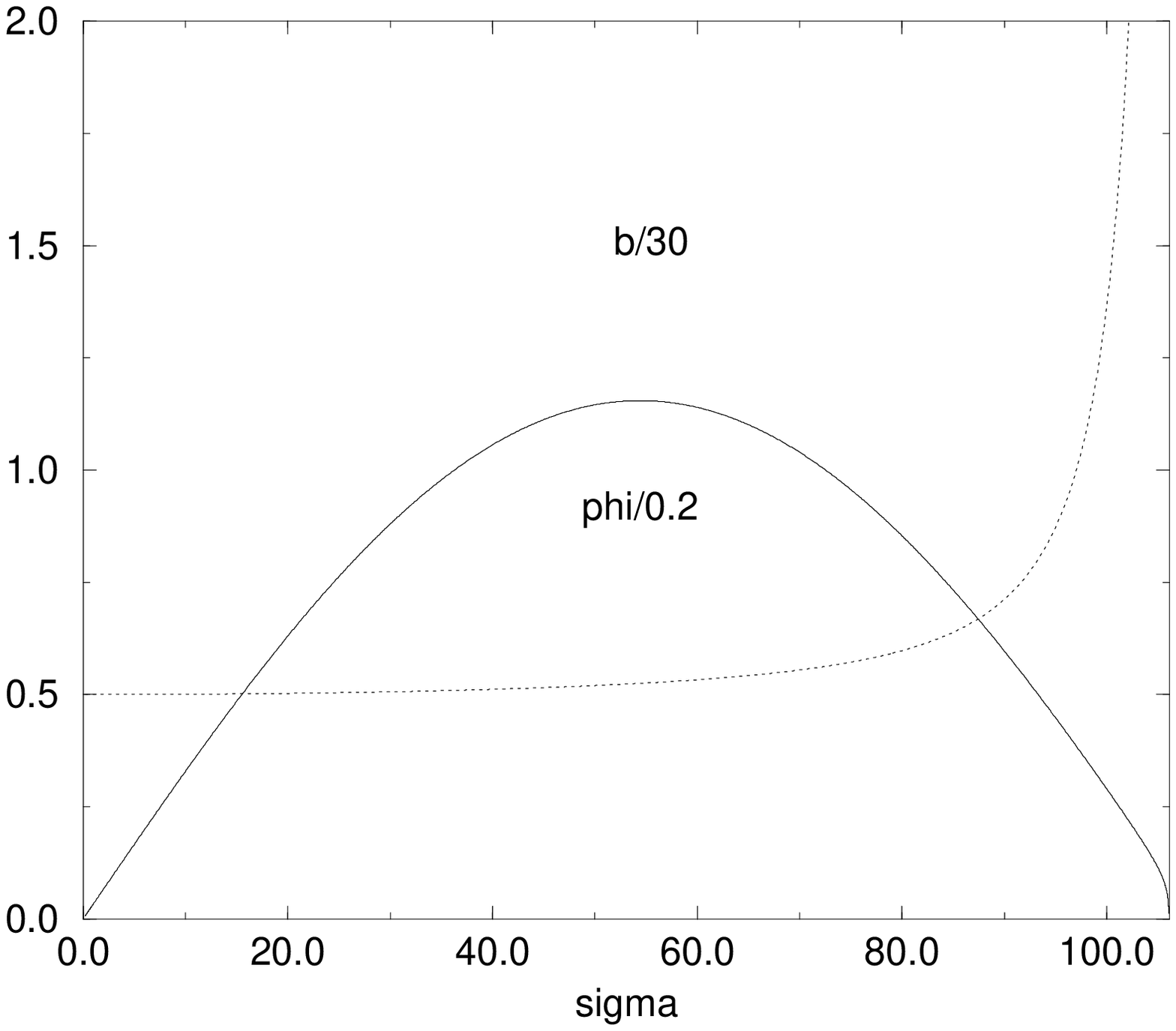}
\caption{
$\varphi(\sigma)$ and $b(\sigma)$ for the HT instanton  
for the potential $V={1\over4}\lambda(1+\varphi^4$) 
with $\lambda=0.01$ and $\varphi_{0}=0.1$.}
\label{fig:caption:bphi1}
\end{figure}

\begin{figure}
\centering
\includegraphics[width=0.50\textwidth]{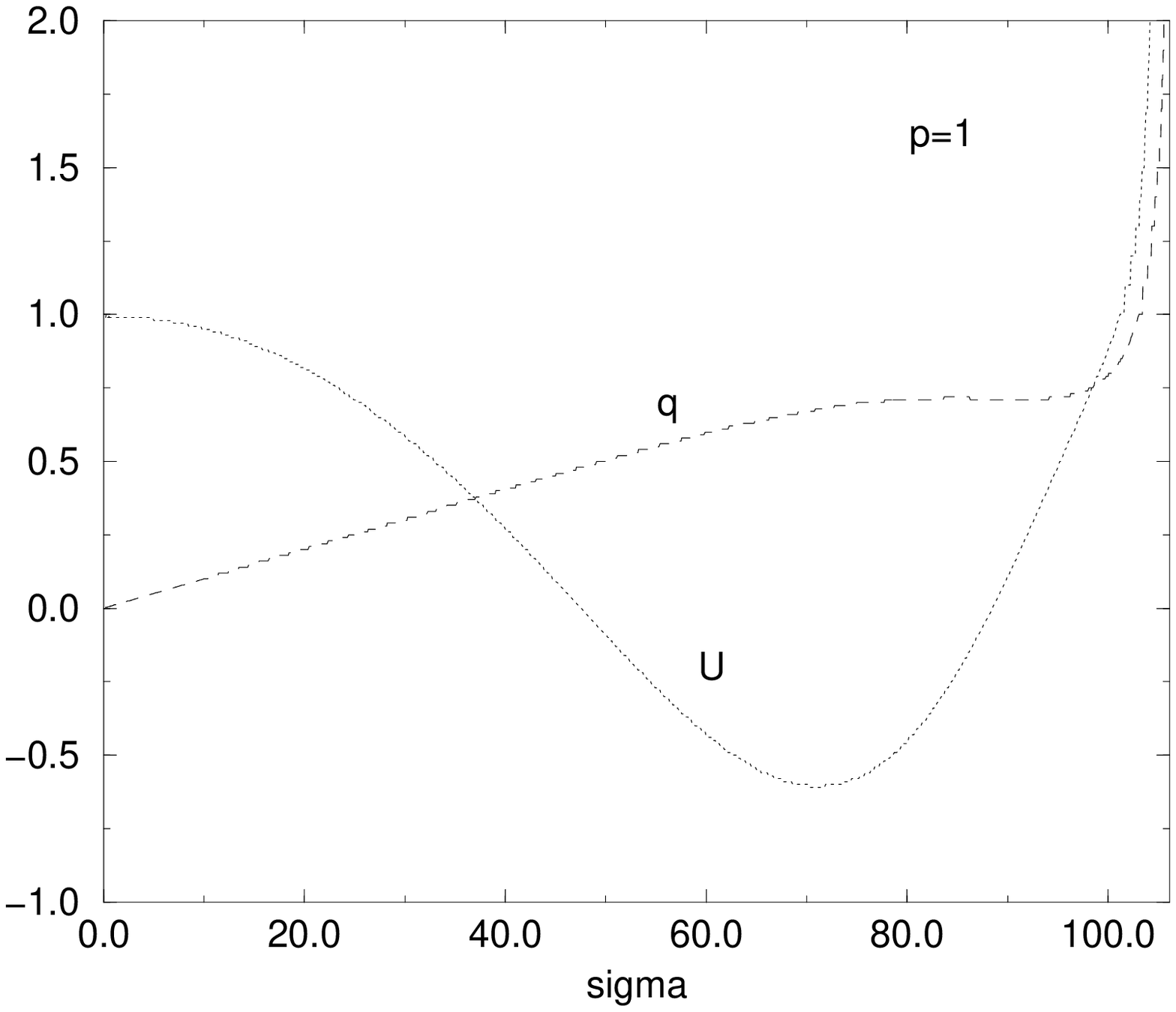}
\caption{
The potential $U$ and the zero energy wave function $q$ 
(for $p=1$ harmonic) 
corresponding to the instanton shown in Fig.~\ref{fig:caption:bphi1}.
The wave function has no nodes.}\label{fig:caption:qU1}
\end{figure}

Typical results for the monotonous potentials are shown 
in Fig.~\ref{fig:caption:bphi1} and Fig.~\ref{fig:caption:qU1}
for the case of a quartic scalar field potential
$V(\varphi)={1\over 4}\lambda (1+\varphi^4)$.
For some initial data the potential $U$ is negative 
in the intermediate region, but in all cases the zero energy wave 
function has no nodes, 
i.e. there are no bound states with negative energies for these potentials.

\begin{figure}
\centering
\includegraphics[width=0.50\textwidth]{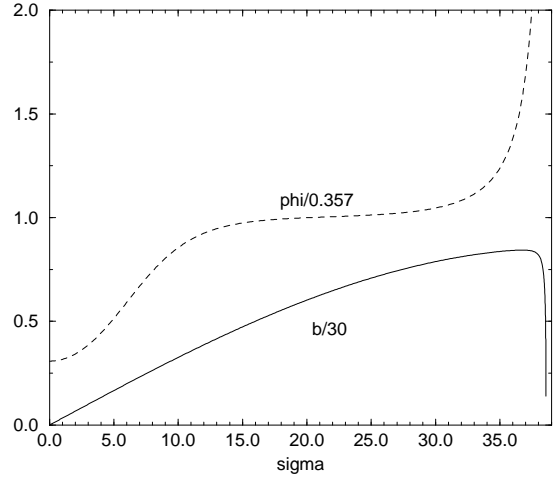}
\caption{
$\varphi(\sigma)$ and $b(\sigma)$ for the HT instanton  
for the potential $V={m^2\over2 }(\varphi^2 (\varphi-v)^2 +B\varphi^4)$  
with $m^2=2, v=0.5, B=0.12$, and $\varphi_{0}=0.11<\varphi_{\star}$.}
\label{fig:caption:bphi3}
\end{figure}

\begin{figure}
\centering
\includegraphics[width=0.50\textwidth]{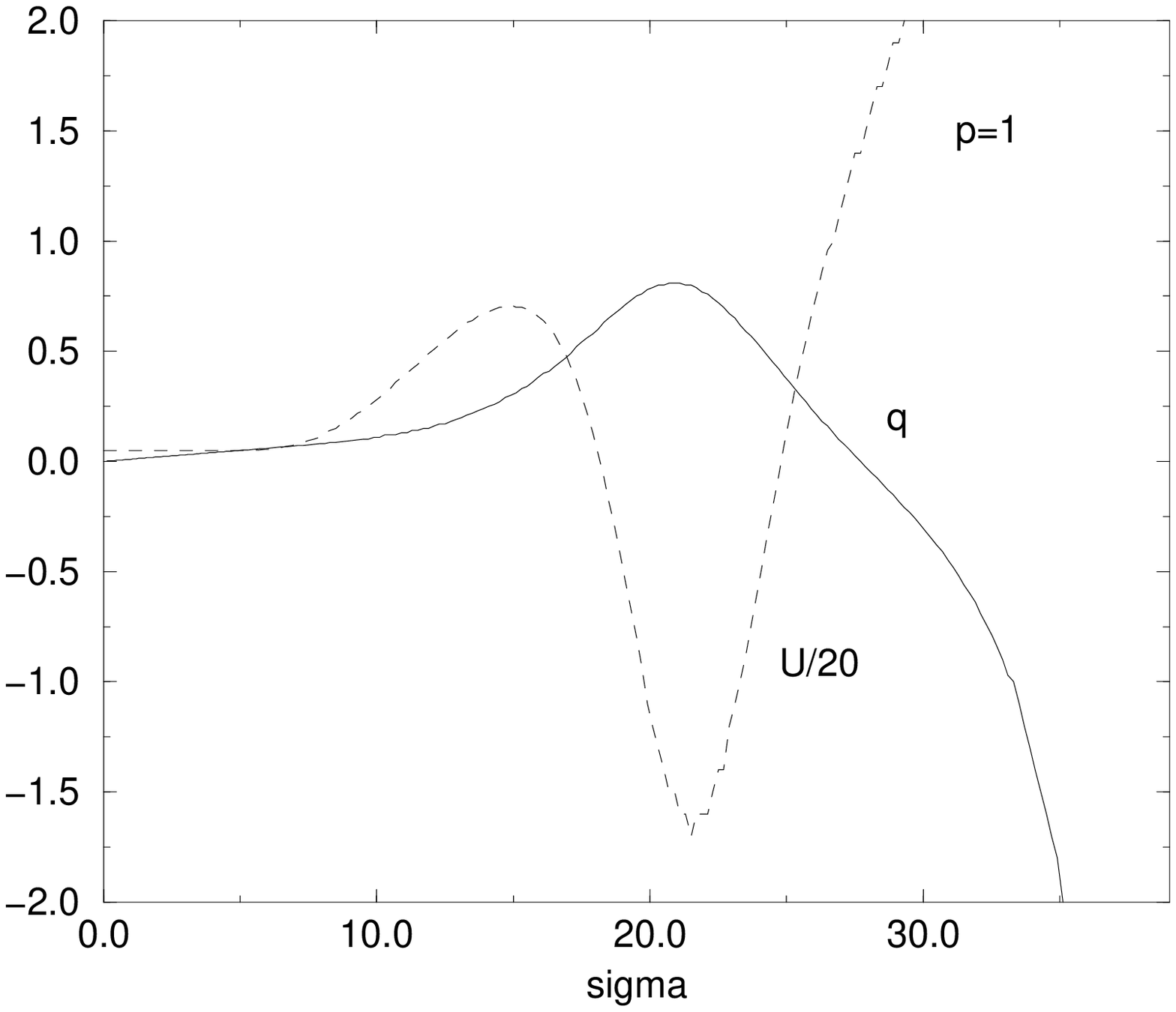}
\caption{
The potential $U$ and the zero energy wave function $q$ 
(for $p=1$ harmonic) 
corresponding to the HT instanton shown in Fig.~\ref{fig:caption:bphi3}.
The wave function has a node.}\label{fig:caption:qU3}
\end{figure}

Next we investigated a potential  
\be
V={m^2\over2 }( \varphi^2 (\varphi-v)^2 +B\varphi^4)  
\ee
with $m^2=2, v=0.5$, and $B=0.12$ (compare \cite{BL}). 
This potential has a local maximum at $\varphi_{top}=0.31250$ and 
a local minimum (false vacuum) at $\varphi_{f}=0.3571429$.
One finds the CD solution for $\varphi_{0}=\varphi_{\star}=0.1123575$.
It was shown in \cite{TS} that there are no negative modes about the 
CD solution.
If one starts integration with  $\varphi_{0}<\varphi_{*}$ one gets 
an ``overshoot'' singular solutions (HT instantons).
We found that for these parameter values HT instantons 
have a negative mode as is shown 
in Fig.~\ref{fig:caption:bphi3} and Fig.~\ref{fig:caption:qU3}.
For the $\varphi_{0}>\varphi_{f}$ the potential is monotonous 
and no negative modes are found about corresponding HT instantons.
For $\varphi_{\star}<\varphi_{0}<\varphi_{f}$ one finds the singular 
solutions or Bousso-Linde double-bubble instantons \cite{BL}.

\section{Concluding remarks}

To conclude,
we demonstrated analytically that the quadratic action for the HT instanton
is non-negative for the harmonics with $p>1$.
For the lowest $p=1$ (homogeneous) harmonic we do not have a general proof.  
We investigated corresponding Schr\"odinger equation numerically
for different scalar field potentials and different scalar field initial 
values and in some cases found a bound state. 
We argue that there is a negative mode about the HT instanton
for certain potentials and certain initial conditions.

In \cite{Wu} it was suggested that the HT solution be considered
as a constrained instanton. 
It is an open question how one can incorporate this idea 
into the stability consideration.
%\footnote{I am thankful to V. Rubakov for this comment.}

As a starting point of our analysis the expression for the quadratic 
action has been taken from the work \cite{GMST},
where it was obtained via Hamiltonian reduction.
It is interesting to note that using the Lagrangian formalism 
one obtains only a sign indefinite quadratic action \cite{LRT,TS},
whereas the Hamiltonian reduction allows one \cite{TS,GMST} to derive 
a quadratic action which has negative overall sign only for the lowest 
(homogeneous) harmonic. 
This problem does not arises for the $K=0$ universe.

Let me mention that the quadratic action in the form Eq.~(\ref{eaction}) 
cannot be straightforwardly used to study the double-bubble instantons of 
Bousso and Linde \cite{BL}, since these solutions have points 
$\dot{\varphi}=0$ in the intermediate region.  

These questions need further investigation.

%\section{Acknowledgments}
\acknowledgments
I am grateful to P. H\'{a}j\'{\i}\v{c}ek, D. Maison, V.A. Rubakov
and T. Tanaka for critical comments, 
and to V. Gusynin for useful discussions.
I would like to thank the Tomalla Foundation and 
the Swiss National Foundation for financial support. 

The main part of the work was done to 
the accompaniment of Celin Dion's music. 
\newpage
%%%%%%%%%%%%%%%%%%%%%%%%%%%%%%%%%%
%%%
%%%%%%%%%%%%%%%%%%%%%%%%%%%%%%%%%%%
%
\end{document}